\documentstyle[preprint,aps,prl]{revtex}
\tightenlines
\begin{document}

%\preprint{{\bf ETH-TH/96-10}}
%\draft

\title{Electrostatics of Vortices in Type II Superconductors}

\author{Gianni Blatter$^{a}$, Mikhail Feigel'man$^{b}$, Vadim
  Geshkenbein$^{a,\, b}$, Anatoli Larkin$^{a,\, b,\, c}$, and Anne van
  Otterlo$^{a}$}

\address{$^{a\,}$Theoretische Physik, ETH-H\"onggerberg, CH-8093
  Z\"urich, Switzerland}

\address{$^{b\,}$L. D. Landau Institute for Theoretical Physics,
  117940 Moscow, Russia}

\address{$^{c\,}$School of Physics and Astronomy, University of
  Minnesota, Minneapolis, MN 55455, USA}

%\twocolumn[
\date{\today}
\maketitle
\widetext
\vspace*{-1.0truecm}
\begin{abstract}
\begin{center}
\parbox{14cm}{

%%%%%%%%%%%%%%%%%%%%%%%%%%%%%%%%%%%%%%%%%%%%%%%%%%%%%%%%%%%%

  In a type II superconductor the gap variation in the core of a
  vortex line induces a local charge modulation. Accounting for
  metallic screening, we determine the line charge of individual
  vortices and calculate the electric field distribution in the half
  space above a field penetrated superconductor. The resulting field
  is that of an atomic size dipole ${\bf d} \sim e a_{{\rm
      \scriptscriptstyle B}} {\bf {\hat z}}$, $a_{{\rm
      \scriptscriptstyle B}} = \hbar^2/m e^2$ is the Bohr radius,
  acting on a force microscope in the pico to femto Newton range.}
\end{center}
\end{abstract}
%]

\narrowtext

The trapping of a magnetic flux $\Phi_\circ = hc/2e$ by a vortex line
in a type II superconductor is a well known phenomenon
\cite{Abrikosov}.  Less familiar, however, is the fact that a vortex
line in general traps an electric charge $Q$ as well. It is the
purpose of this letter to determine this vortex line charge
quantitatively and to discuss the feasibility of its experimental
observation.

The vortex line charge $Q$ has been discussed before by Khomskii and
Freimuth \cite{Khomskii} and by Feigel'man {\it et al.}
\cite{Feigelman} (see also \cite{Otterlo}) within the context of the
sign change in the Hall coefficient, as observed in a number of type
II superconductors \cite{Hagen}. Here, we concentrate on the vortex
charge and its accompanying electrostatic features, with a specific
emphasis on its experimental observability.

The main reason for the charge accumulation around the vortex is found
in the particle-hole asymmetry, as quantified by the energy dependence
of the density of states (DOS, per spin) $N(E)$ at the Fermi level, $Q
\propto d N(E)/d E |_\mu$. In the presence of particle-hole asymmetry,
the carrier density $n(\mu, \Delta)$ in the superconductor not only
depends on the chemical potential $\mu$, but on the energy gap
$\Delta$ as well. The singular behavior of the phase at the center of
a vortex leads to the formation of a core with a suppressed gap
function $\Delta(R \rightarrow 0) \rightarrow 0$, where $R$ measures
the distance from the phase singularity.  With the chemical potential
fixed, charge carriers (electrons/holes for a nearly filled/empty
band) are expelled from this core region.  Metallic screening
drastically reduces the accumulated charge, however, in our analysis
below, we show that the residual vortex line charge $|Q| \sim e
k_{{\rm\scriptscriptstyle F}} (\lambda_{{\rm\scriptscriptstyle
    TF}}/\xi)^2$ is still experimentally observable (here,
$k_{{\rm\scriptscriptstyle F}}$, $\lambda_{{\rm\scriptscriptstyle
    TF}}$, and $\xi$ denote the Fermi wave number, the Thomas-Fermi
screening length, and the coherence length, respectively; we define
$e>0$). In particular, one may envisage the classic geometry for the
observation of vortices via the Bitter-decoration method
\cite{TraubleEssmann}, with the superconductor filling the half-space
$z < 0$ and penetrated by a magnetic field ${\bf B} \parallel {\bf
  {\hat z}}$, see Fig.\ 1. The vortex line-charge produces an electric
field in the vacuum above the superconductor ($z > 0$), which
corresponds to the one of a surface electric dipole ${\bf d} \parallel
{\bf {\hat z}}$ with unit $\pm$ charges separated by a distance $\sim
1$ \AA.  The vortex charge is associated with the core size $\xi$ and
therefore a much higher resolution can be expected in an electrostatic
experiment as compared to the magnetic experiments probing structures
on the scale of the penetration depth $\lambda_{\rm \scriptscriptstyle
  L}$.  In the following, we derive the vortex line charge $Q$ and
solve the ``half-space" electrostatic problem for a single vortex and
for the vortex lattice. Next, we discuss the observability of the
vortex charge and close with a few more subtle questions regarding our
analysis.

The origin of the vortex charge can be understood on the basis of a
textbook problem \cite{Ashcroft}: Consider the Sommerfeld free
electron model for a metal and determine the particle density $n$ at
{\it fixed chemical potential} $\mu$, $n(T) \approx n(0)
[1+(\pi^2/8)(T^2/\mu^2)]$. The density increase $\delta n$ is a
consequence of the finite temperature smearing of the Fermi function
combined with the finite slope in the density of states $N'_\mu \equiv
d N(E) / d E |_\mu$ ($= 3 n /8 \mu^2$ in a 3D parabolic band). For a
general Fermi surface with a smooth DOS we have
\begin{eqnarray}
  \delta n \approx (\pi T)^2 N'_\mu/3.
\label{dnM}
\end{eqnarray}
Note that the sign of $\delta n$ depends on that of $N'_\mu$, with
$\delta n > 0$ for electron like carriers \cite{remark}.

Next, consider a BCS superconductor where the pair occupation
probability $v_{\bf k}^2 = (1-\xi_{\bf k}/E_{\bf k})/2$ ($\xi_{\bf k}
= \epsilon_{\bf k} - \mu$ and $E_{\bf k} = (\xi_{\bf k}^2
+\Delta^2)^{1/2}$ denote the excitation energies in the normal and
superconducting state, respectively) determines the density via $n = 2
\sum_{\bf k} v_{\bf k}^2$. The opening of a gap $\Delta$ in the
spectrum produces an analogous smearing in the occupation probability
and the density changes with $\Delta$ according to
\begin{eqnarray}
  \delta n \approx \Delta^2 N'_\mu \ln(\hbar\omega_{{\rm
      \scriptscriptstyle D}}/T_c),
\label{dnBCS}
\end{eqnarray}
where $\omega_{{\rm \scriptscriptstyle D}}$ denotes the usual
frequency cutoff on the attractive interaction.  Indeed, the
occupation probability in the superconductor resembles a Fermi
distribution with a temperature $T \simeq \Delta$ \cite{Tinkham}, in
agreement with the results (\ref{dnM}) and (\ref{dnBCS}).

In the presence of a vortex line the gap parameter turns to zero in
the core,
\begin{eqnarray}
  \Delta^2 (R) \approx \Delta^2_\infty \, R^2/(R^2+\xi^2),
\label{Deltar}
\end{eqnarray}
where $R < \lambda_{{\rm \scriptscriptstyle L}}$ denotes the radial
distance from the phase singularity, $\Delta_\infty$ is the magnitude
of the gap parameter far away from the core, and the coherence length
$\xi$ determines its spatial extent. The slow algebraic decay $\delta
\Delta^2 \sim - \Delta^2_\infty \, (\xi/R)^2$ is a consequence of the
slow decay of the supercurrent $j(R) \sim j_\circ \xi/R$ within the
London screening length $\lambda_{{\rm \scriptscriptstyle L}}$
($j_\circ$ is the depairing current density; $\delta \Delta^2$ drops
to zero exponentially for $R > \lambda_{{\rm \scriptscriptstyle L}}$).

We account for metallic screening within a Thomas-Fermi approximation,
substituting $\mu$ by the electrochemical potential $\mu + e \varphi$
in the expression for the density $n$. The density modulation $\delta
n(R) = n[\mu +e \varphi(R), \Delta(R)] -n(\mu,\Delta_\infty)$ is
driven by the variation in the gap function $\Delta(R)$ and induces a
scalar potential $\varphi(R)$, which is obtained from the solution of
Poisson's equation
\begin{eqnarray}
  \nabla^2 \varphi (R) = 4 \pi e \delta n(R).
\label{Poisson}
\end{eqnarray}
Linearizing in $\varphi$ and $\Delta$ we arrive at 
\begin{eqnarray}
  [\nabla^2 - \lambda^{-2}_{{\rm \scriptscriptstyle TF}}] \varphi (R)
  = 4 \pi e \delta n_{\rm ext} (R),
\label{scPoisson}
\end{eqnarray}
with the Thomas-Fermi length $\lambda_{{\rm \scriptscriptstyle TF}} =
(8 \pi e^2 N_\mu)^{-1/2}$ and the ``external" density modulation
\begin{eqnarray}
  \delta n_{\rm ext}(R) = - N_\mu \Delta_\infty^2 \frac{\xi^2}{R^2
    +\xi^2} \frac{d \ln T_c}{d \mu}
\label{dngen}
\end{eqnarray}
(we have substituted the expression $N'_\mu \ln(\hbar \omega_{{\rm
    \scriptscriptstyle D}}/T_c)$, see (\ref{dnBCS}), by the
phenomenological parameter $N_\mu d \ln T_c /d \mu$ using the BCS
expression $T_c \approx \hbar \omega_{{\rm \scriptscriptstyle D}}
\exp(-1/N_\mu V)$).  The integration of (\ref{dngen}) over the planar
coordinate ${\bf R}$ provides the total external line charge
\begin{eqnarray}
  Q_{\rm ext} \approx 2 \pi e \Delta_\infty^2 \xi^2 N_\mu \frac{d \ln
    T_c}{d \mu} \ln \frac{\lambda_{{\rm \scriptscriptstyle L}}}{\xi}.
\label{rhoext}
\end{eqnarray}
For a BCS model in the clean limit $\Delta_\infty^2 \xi^2 N_\mu =
k_{{\rm \scriptscriptstyle F}} {\tilde \mu}/\pi^4$ with ${\tilde \mu}
= m_{{\rm eff}} v^2_{{\rm \scriptscriptstyle F}}/2$ (for a nontrivial
Fermi surface we have $\mu \neq {\tilde \mu}$ in general), and using
$d \ln T_c / d \ln {\tilde \mu} \approx \ln (\hbar \omega_{{\rm
    \scriptscriptstyle D}} / T_c) \sim 1$ -- 10, we obtain an external
line charge of order $e k_{{\rm \scriptscriptstyle F}}$, with only a
weak dependence on the superconducting parameters $T_c$ and
$\lambda_{{\rm \scriptscriptstyle L}}$.

We determine the real charge distribution $\rho(R) = -e \delta n(R) =
-\nabla^2 \varphi(R)/4\pi$ (positive for electron-like carriers) by
solving the screened Poisson equation (\ref{scPoisson}). In the limit
$\lambda_{{\rm \scriptscriptstyle TF}} \ll \xi$ we obtain
\begin{eqnarray}
  \rho(R) = \frac{e a_{{\rm \scriptscriptstyle B}}}{\pi^3} \frac{d\ln
    T_c}{d\ln {\tilde \mu}} \frac{\xi^2 - R^2}{(R^2 +\xi^2)^3},
\label{charge}
\end{eqnarray}
where $a_{{\rm \scriptscriptstyle B}}$ denotes the Bohr radius.
Overall charge neutrality requires the total charge $\int d^2R
\rho(R)$ to vanish: The line charge accumulated within the vortex core
is $Q_\xi = e a_{{\rm \scriptscriptstyle B}}(d\ln T_c/d\ln {\tilde
  \mu})/(4 \pi \xi)^2 \approx Q_{\rm ext} \lambda^2_{{\rm
    \scriptscriptstyle TF}}/\xi^2$ and an equal and opposite charge is
provided by the screening outside the core region, see Fig.\ 1.

Next, we solve the electrostatic problem for a single vortex line
penetrating a superconductor filling the lower half space $z<0$, see
Fig.\ 1. The potential $\varphi(R,z)$ generated by the charge density
$\rho_{\rm ext} (R,z)$ $= -e \delta n_{\rm ext}(R) \Theta(-z)$ is
obtained by solving the (screened) Poisson equation $[\nabla^2
-\lambda_{{\rm \scriptscriptstyle TF}}^{-2} \Theta(-z)]\varphi(R,z) =
-4 \pi \rho_{\rm ext} (R,z)$. We decompose the potential into a bulk
and an interface term, $\varphi(R,z) = \varphi_\infty(R) \Theta(-z) +
\varphi_0(R,z)$, where $\varphi_\infty(R)$ denotes the bulk solution
of (\ref{scPoisson}).  The interface term $\varphi_0(R,z)$ can be
obtained from the Fourier Ansatz
\begin{eqnarray}
  \varphi_0(R,z) = \int \frac{d^2K}{(2\pi)^2} \varphi_0^\pm ({\bf K})
  \exp[i{\bf K}{\bf R} \mp k_z^\pm z],
\label{varphi0}
\end{eqnarray}
with $k_z^+ = K$ and $k_z^- = \sqrt{K^2+\lambda_{{\rm
      \scriptscriptstyle TF}}^{-2}}$ referring to values $z > 0$ above
and $z < 0$ below the vacuum -- superconductor interface. Requiring
continuity of $\varphi$ and $\nabla \varphi$ across the interface, we
can express $\varphi_0^\pm({\bf K})$ through the source term
$\varphi_\infty(K) = 4 \pi \rho_{{\rm ext}} (K)/(K^2+\lambda_{{\rm
    \scriptscriptstyle TF}}^{-2})$. After transformation back to real
space we arrive at the final expression
\begin{eqnarray}
  \varphi(R,z>0) &=& \int \frac{d^2K}{(2\pi)^2} \varphi_\infty(K)
  \frac{\sqrt{K^2+\lambda_{{\rm \scriptscriptstyle TF}}^{-2}}}
  {K+\sqrt{K^2+\lambda_{{\rm \scriptscriptstyle TF}}^{-2}}}
  \nonumber\\ &\phantom{.}& \qquad\qquad\times\exp(i{\bf K}{\bf R} -
  Kz).
\label{varphi}
\end{eqnarray}
The integral is dominated by small wave vectors and we may neglect
$K^2$ as compared to $\lambda_{{\rm \scriptscriptstyle TF}}^{-2}$.
Using $\rho_{{\rm ext}} (K) \approx Q_{{\rm ext}}
K_0(K\xi)/\ln(\lambda_{{\rm \scriptscriptstyle L}}/\xi)$, with $K_0$
the modified Bessel function, the integration over ${\bf K}$ yields
\begin{eqnarray}
  e \varphi(R,z) [{\rm eV}] &\approx& 0.8 \frac{m}{m_{\rm eff}}
  \frac{d \ln T_c}{d \ln {\tilde \mu}} \ln\frac{\min(z,\lambda_{{\rm
        \scriptscriptstyle L}})}{\xi}\nonumber\\ 
  &\phantom{.}&\qquad\qquad \times \frac{z[{\rm
      \AA}]}{(R^2+z^2)^{3/2}},
\label{potential}
\end{eqnarray}
where we have chosen $z > \xi$ and all lengths are taken in
{\AA}ngstr\"oms (note that (\ref{potential}) is independent of the
DOS, the latter appearing both in $\delta n_{\rm ext}$ and in the
screening length $\lambda_{{\rm \scriptscriptstyle TF}}^{2}$, and only
weakly depends on the superconducting properties). The result
(\ref{potential}) is the potential generated by a surface dipole ${\bf
  d}\parallel {\bf {\hat z}}$ smeared on the scale $\xi$,
$\varphi({\bf r}) = {\bf d}\cdot{\bf r} /r^3$,
\begin{eqnarray}
  {\bf d} = \frac{e a_{{\rm \scriptscriptstyle B}} {\bf {\hat
        z}}}{\pi^2}\frac{m}{m_{\rm eff}} \frac{d\ln T_c}{d\ln {\tilde
      \mu}}\ln\frac{\min(z,\lambda_{{\rm \scriptscriptstyle
        L}})}{\xi}.
\label{dipole}
\end{eqnarray}
With the logarithms roughly compensating for the numerical $\pi^{-2}$,
we find a dipole with unit $\pm$ charges separated by $\sim 1$ \AA.
The charge and field geometry are illustrated in Fig.\ 1.

The corresponding electrostatic problem for a vortex lattice is solved
in the same manner. The integral $\int d^2K \exp[i{\bf K}{\bf R} -
Kz]$ producing the dipole field has to be replaced by the sum over
reciprocal lattice vectors ${\bf K}_{\bf n}$ of the vortex lattice,
$(2\pi/a_\triangle)^2 \sum_{\bf n} \exp[i{\bf K}_{\bf n}{\bf R} -
K_{\bf n}z]$, where $a_\triangle = (2/\sqrt{3})^{1/2}
(\Phi_\circ/B)^{1/2}$ denotes the lattice constant. Usually the sum
can be restricted to the six nearest neighbor lattice vectors and the
result reads
\begin{eqnarray}
  e \varphi_{{\rm \scriptscriptstyle VL}} (R,z) [{\rm eV}] &\approx&
  15.0 \frac{m}{m_{\rm eff}} \frac{a_{{\rm \scriptscriptstyle
        B}}^2}{a_\triangle^2} \frac{d \ln T_c}{d \ln {\tilde \mu}}
  \ln\frac{\min(z,a_\triangle,\lambda_{{\rm \scriptscriptstyle
        L}})}{\xi}\nonumber\\ &\times& \left[1+\exp(-2\pi
    z/a_\triangle) \sum_{\rm n.n.} \cos {\bf K}_{\bf n}{\bf R}\right].
\label{lattice}
\end{eqnarray}
This completes our derivation of the charge- and the electrostatic
field distribution for the individual vortex and the vortex lattice.

Is this vortex charge observable in an experiment? The most
straightforward attempt to identify the vortex charge is based on
(scanning) force microscopy.  Indeed, the observation of single charge
carriers by force microscopy has been reported by Sch\"onenberger and
Alvarado \cite{Schonenberger}. Below we consider two experimental
setups: i) A grounded metallic tip is approached to an individual
vortex. The vortex (surface) dipole induces a second dipole in the
tip, resulting in a dipole -- dipole attraction between the vortex and
the tip.  The expected force is estimated to be of the order of $F_{\,
  \rm ind} \sim 10^{-17}$ N.  ii) The metallic tip is biased against
the superconductor. In this capacitor geometry, the bias voltage $V$
drives a charge transfer from the superconductor to the tip, leading
to a tip -- surface attraction $F_{\rm ts}$ which has to be
compensated in the experiment. As the tip is approached to the vortex,
the vortex-dipole -- tip-charge interaction produces an additional
force on the tip, which is the desired signal. The estimated force is
proportional to the bias voltage $V$ and is of the order of $F_{\, \rm
  bias} \sim 10^{-14}\, V[V]$ N.  Note, that the mobile vortex dipole
can be distinguished from static charged surface defects, as produced
by adsorbed atoms and molecules, by driving the vortex with an
external ac force and using a lock-in technique.

In order to estimate the dipole-dipole attraction in the first setup
i) we model the tip as a metallic sphere of radius $\rho$. Its center
is chosen a distance $\zeta > \rho$ right above the vortex, thus
producing the maximal tip -- vortex attraction. A straightforward
calculation using the image charge technique (see, e.\ g., Ref.
\cite{Jackson}; we ignore higher order images) provides the result
\begin{eqnarray}
  F_{\, \rm ind} = 2\rho\zeta
  d^2\frac{2\rho^2+\zeta^2}{(\zeta^2-\rho^2)^4}.
\label{Find}
\end{eqnarray}
Inserting the expression (\ref{dipole}) for the vortex-dipole, using
$e^2/a_{{\rm \scriptscriptstyle B}}^2 = 8.23 \cdot 10^{-8}$ N, and
choosing a typical geometry with $\rho \sim \zeta/2$, we find
\begin{eqnarray}
  F_{\, \rm ind} \sim 2.5 \cdot 10^{-8} \left(\frac{a_{{\rm
          \scriptscriptstyle B}}}{\rho}\right)^4 \, {\rm N},
\label{Findn}
\end{eqnarray}
where we have assumed that $(m/m_{\rm eff})(d\ln T_c/ d\ln {\tilde
  \mu})$ $\ln[\min(\zeta,\lambda_{{\rm \scriptscriptstyle L}})/\xi]
\sim 10$.  With $\rho \sim 100$ {\AA} the resulting force is $F_{\,
  \rm ind} \sim 2 \cdot 10^{-17}$ N.

Next, we consider the capacitor setup ii) where the tip is biased
against the superconductor. It is convenient to model this geometry
with a tip of spheroidal shape. Using elliptic coordinates (see, e.\ 
g., Ref. \cite{Abramowitz}), we define the superconductor and tip
surfaces through the coordinates $\eta_{\rm sc} = 0$ and $\eta_{\,\rm
  tip} = \eta_\circ$, respectively.  Of the three parameters, the
surface -- tip distance $\zeta$, the tip radius of curvature $\rho$,
and the tip aperture $2 \vartheta$, only two can be freely chosen,
$\rho/\zeta = \tan^2 \vartheta = (1-\eta_\circ^2)/\eta_\circ^2$.
Solving Poisson's equation $\Delta V = 0$, we find $V(\eta) = V
g(\eta)/ g(\eta_\circ)$, resulting in an electric field ${\bf E}_{{\rm
    tip}}(a,R) = - 2 V {\bf {\hat z}}/ g(\eta_\circ) (a^2 +
R^2)^{1/2}$ at the superconductor -- vacuum interface.  Here, $g(\eta)
= \ln(1+\eta)/(1-\eta)$ and $a = \zeta/\eta_\circ$ is the scale factor
in the transformation to elliptic coordinates. The energy of the
vortex-dipole in the electric field of the biased tip is given by
$U(a,R) = - {\bf d}\cdot {\bf E}_{{\rm tip}}(a,R)/2$ (only half-space)
and for the forces perpendicular and parallel to the surface we obtain
\begin{eqnarray}
  (F_{\, \rm bias}^\perp, F_{\, \rm bias}^\parallel) = \frac{d
    V}{g(\eta_\circ)} \left( \frac{1}{a^2}, \frac{1}{R^2
      (1+a^2/R^2)^{3/2}} \right)
\label{Fbias}
\end{eqnarray}
(see Ref. \cite{Schonenberger} for the description of an ac technique
used to separate the small modulation $F_{\, \rm bias}^\perp$ due to
the vortex-dipole from the large base force $F_{\rm ts}$ due to the
image charge attracting the tip).  The above results apply for
distances $> \xi$; upon approaching the tip closer to the vortex the
details of the charge distribution become relevant and the forces
change, e.\ g., in $F_{\, \rm bias}^\perp$ we have to replace the
scale $a$ by the spread of the vortex charge $\xi$.  Using $\zeta \sim
\rho$ and $\vartheta = \pi/4$ we obtain the numerical expression
\begin{eqnarray}
  F_{\, \rm bias} \sim 10^{-9}\, V [{\rm V}] \left(\frac{a_{{\rm
          \scriptscriptstyle B}}}{\rho}\right)^2 {\rm N}.
\label{Fbiasn}
\end{eqnarray}
With $\rho \sim 100$ {\AA} the force amounts to $F_{\, \rm bias} \sim
3 \cdot 10^{-14}\, V[{\rm V}]$ N. At present, forces in the pN range
are observed in state of the art AFM experiments and the fN regime is
being attacked in the near future.

A number of alternative experiments detecting the vortex line charge
$Q$ look promising as well. Here we only mention the basic ideas. i)
One of the most sensitive electrometers is the
single-electron-transistor (SET), e.\ g., see Ref.\ \cite{SET}, where
the small central island is connected via tunnel junctions to the two
leads. A capacitively coupled gate takes the device to its optimal
working point. The device has to be fabricated onto the superconductor
surface with only a thin insulating layer ($\sim 100$ \AA) decoupling
the two systems electronically. Vortices driven across the central
island act to modulate the gate voltage via their line charge and the
signal can be picked up via a lock-in technique.  Using the setup i)
above we can estimate the induced charge on the island to be $\sim d
\rho/\zeta^2 \sim 10^{-2} e$, which should be well detectable by an
SET with a charge resolution of $\sim 10^{-4} e /\sqrt{{\rm Hz}}$
\cite{SET}.

Another straightforward idea is to imitate the original decoration
technique of Tr\"auble and Essmann\cite{TraubleEssmann}, using
electric rather than magnetic particles. It seems difficult, however,
to imagine an electric analogue of the small ferromagnetic (Fe, Co, or
Ni) particles, being assembled and spread onto the superconductor
surface in a similar fashion. Alternatively, one may resort to the use
of electrons, separated from the superconductor surface by a thin
$^4$He film, and setting up a Wigner crystal. The interaction between
the surface dipole array due to the vortex lattice with the 2D Wigner
crystal will lead to new features affecting the physics of the vortex
lattice and the Wigner crystal as well.

We turn to some more subtle issues. On a phenomenological level we can
express the external density modulation (\ref{dngen}) through the
Ginzburg-Landau energy density ${\cal F} = N_\mu [\alpha |\Delta|^2 +
\frac{\beta}{2} |\Delta|^4 + \gamma |\nabla \Delta|^2]$, with the
parameters $\alpha$, $\beta$, and $\gamma$ depending on the chemical
potential $\mu$, $\delta n_{\rm ext}(R)= - \partial_\mu {\cal F}
|_R^\infty$.  The term $N_\mu (\partial_\mu \alpha) |\Delta|^2 \propto
N'_\mu \ln(\hbar\omega_{{\rm \scriptscriptstyle D}}/T_c)(1 - T/T_c)$
produces the main contribution due to particle-hole asymmetry. A
second term $\propto N'_\mu (1 - T/T_c)^2$ originates from taking the
derivative of the prefactor $N_\mu$ in ${\cal F}$. Finally, a third
contribution from the gradient term, $(N_\mu/\mu) (1-T/T_c)
|\xi\nabla\Delta|^2 \propto (N_\mu/\mu)(1-T/T_c)^2$, is present even
without particle-hole asymmetry and produces the well known London
electric field ${\bf E}_{{\rm\scriptscriptstyle L}} \propto \nabla
v_s^2(R)$, compensating the centripetal force of the rotating
superfluid. In our analysis above we have concentrated on the leading
contribution in $(1-T/T_c)$ and in $\ln(\hbar\omega_{{\rm
    \scriptscriptstyle D}}/T_c)$. Away from $T_c$ the other terms will
slightly modify our result.

In summary, we have determined the line charge associated with the
formation of a vortex in a type II superconductor. The charge is
mainly driven by the particle-hole asymmetry, its bare value is
$Q_{{\rm ext}} \sim e k_{{\rm \scriptscriptstyle F}}$, and screening
reduces this value down to $Q \sim e k_{{\rm \scriptscriptstyle F}}
\lambda^2_{{\rm \scriptscriptstyle TF}}/ \xi^2$. We have solved the
electrostatic problem related to the observation of the vortex charge
on a superconductor surface and found the associated electric field to
be that of a microscopic dipole ${\bf d} \sim e a_{{\rm
    \scriptscriptstyle B}} {\bf {\hat z}}$. With our results we hope
to motivate new experiments looking for the vortex charge itself (with
an interesting relation to the sign change of the Hall effect), we
propose a new imaging technique able to address the structure of
vortex systems on the scale $\xi$, and we suggest to use the charge
array set up by the vortices in other experimental areas such as the
problem of electrons on $^4$He films.

We thank J. Aarts, D. Est\`eve, L. Ioffe, Y. Iye, P. Kes, V. Kravtsov,
M. Marchevsky, J. Rhyner, W. van Saarloos, C. Sch\"onenberger, and A.
Volodin for helpful discussions. Financial support from the Swiss
National Foundation and from RFBR (\# 95-02-05720) is gratefully
acknowledged.

\begin{figure}
\makebox[1.75in]{\rule[1.125in]{0in}{1.125in}}
\includegraphics{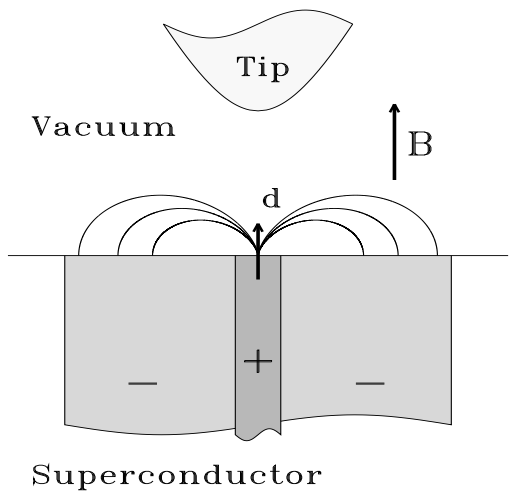}
\vglue 1.0truecm
\caption{The superconductor (lower half-space) is penetrated
  by the magnetic field ${\bf B}$. The resulting vortex line is
  charged due to the particle-hole asymmetry as quantified by the
  finite derivative of the density of states at the Fermi level, $d
  N(E)/dE |_\mu$. Charge carriers are expelled from the vortex core
  (core radius $\sim \xi$, $+$ region in the figure) and an equal and
  opposite screening charge on a scale $\min(a_\triangle,
  \lambda_{{\rm\scriptscriptstyle L}})$ accounts for charge
  neutrality. The electric field generated in the upper half space
  (see field lines in the figure) is that of a surface dipole {\bf d}
  of size $\sim e a_{\rm \scriptscriptstyle B}$.  A tip approaching
  the surface is attracted to the dipole with a force depending on the
  specific setup, see Eqs.\ (15) for a grounded tip and (17) for a tip
  biased with a voltage $V$ against the superconductor.  }
\label{fig:1}
\end{figure}


\begin{thebibliography}{99}

\bibitem{Abrikosov} A.\ A.\ Abrikosov, Zh.\ Eksp.\ Teor.\ Fiz.\ {\bf
        32}, 1442 (1957) [Sov.\ Phys.\ JETP {\bf 5}, 1174 (1957)].
\bibitem{Khomskii} D.\ I.\ Khomskii and A.\ Freimuth,
        {Phys.\ Rev.\ Lett.\ } {\bf 75}, 1384 (1995).
\bibitem{Feigelman} M.\ V.\ Feigel'man {\it et al.}, 
        Pis'ma Zh.\ Eksp.\ Theor.\ Fiz.\ {\bf 62}, 
        811 (1995) [JETP Lett.\ {\bf 62}, 834 (1995)].
\bibitem{Otterlo} A.\ van Otterlo {\it et al.}, 
        {Phys.\ Rev.\ Lett.\ } {\bf 75}, 3736 (1995).
\bibitem{Hagen} S.\ J.\ Hagen {\it et. al.}, 
        {Phys. Rev. B} {\bf 47}, 1064 (1993).
\bibitem{TraubleEssmann} H.\ Tr\"auble and U.\ Essmann, {Phys.\ 
        Stat.\ Sol.\ } {\bf 18}, 813 (1966) and {J.\ Appl.\ Phys.\ }
        {\bf 39}, 4052 (1968).
\bibitem{Ashcroft} N.\ W.\ Ashcroft and N.\ D.\ Mermin, 
        {\it Solid State Physics} 
        (Saunders College HRW, Philadelphia, 1976).
\bibitem{remark} The correspondence between the sign of $N_\mu'$ 
        and that of the charge carriers is somewhat subtle and 
        can be reversed, particularly in reduced dimensions, see 
        \cite{Otterlo}.
\bibitem{Tinkham} M.\ Tinkham, {\it Introduction to
        Superconductivity} (Krieger, 1980).  
\bibitem{Schonenberger} C.\ Sch\"onenberger and S.\ F.\ Alvarado, 
        {Phys.\ Rev.\ Lett.\ } {\bf 65}, 3162 (1990).
\bibitem{Jackson} J.\ D.\ Jackson, 
        {\it Classical Electrodynamics} (Wiley, 1962).
\bibitem{Abramowitz} M.\ Abramowitz and I.\ E.\ Stegun, 
        {\it Handbook of Mathematical Functions} 
        (Dover, New York, 1965).
\bibitem{SET} A.\ N.\ Cleland {\it et al.}, 
         Appl.\ Phys.\ Lett.\ {\bf 61}, 2820 (1992).
\end{thebibliography}
\end{document}